# CONSISTENT PRICE SYSTEMS AND FACE-LIFTING PRICING UNDER TRANSACTION COSTS

By Paolo Guasoni,[1] Miklós Rásonyi[2,3]
and Walter Schachermayer[3]

*Boston University, Computer and Automation Institute of the Hungarian Academy of Sciences and Vienna University of Technology and Vienna University of Technology*

In markets with transaction costs, consistent price systems play the same role as martingale measures in frictionless markets. We prove that if a continuous price process has conditional full support, then it admits consistent price systems for arbitrarily small transaction costs. This result applies to a large class of Markovian and non-Markovian models, including geometric fractional Brownian motion.

Using the constructed price systems, we show, under very general assumptions, the following "face-lifting" result: the asymptotic superreplication price of a European contingent claim $g(S_T)$ equals $\hat{g}(S_0)$, where $\hat{g}$ is the concave envelope of $g$ and $S_t$ is the price of the asset at time $t$. This theorem generalizes similar results obtained for diffusion processes to processes with conditional full support.

**1. Introduction.** It is common wisdom in mathematical finance that simple and elegant statements in frictionless markets are torn apart and made almost unrecognizable by the slightest market imperfection.

Transaction costs are arguably the most pervasive of frictions as they affect virtually all financial (and also nonfinancial) markets, and the presence

---
Received February 2007; revised June 2007.
[1]Supported by NSF Grant DMS-05-3239.
[2]Supported by Hungarian National Science Foundation (OTKA) Grants F 049094 and T 047193.
[3]Supported by Austrian Science Foundation (FWF) Grant P 19456, from Vienna Science and Technology Fund (WWTF) Grant MA 13 and from the Christian Doppler Research Association (CDG).

of bid and ask prices is crucial in the equilibrium between traders and market makers, which ensures market liquidity.

Since bid–ask spreads are usually a tiny fraction of prices, the theory of frictionless markets rests on the tacit assumption that small transaction costs have small effects. It is then unsettling that the basic questions of no-arbitrage and superreplication have radically different solutions with transaction costs, which moreover do *not* converge to the frictionless solution as the bid–ask spread shrinks to a single price [7, 10, 20, 28, 40].

In a frictionless market, the "no free lunch with vanishing risk" (NFLVR) condition for simple strategies dictates that locally bounded asset price processes are semimartingales and (NFLVR) for general strategies further implies the existence of a local martingale measure ([14], Theorems 7.1 and 1.1). The superreplication price of a contingent claim is then obtained as the supremum of its expected value under all such measures ([14], Theorem 5.7). When this measure is unique, the market is complete and the price of a claim is uniquely determined.

By contrast, with transaction costs, arbitrage disappears for several examples of nonsemimartingales, such as fractional Brownian motion [17]. Furthermore, market completeness fails except in trivial cases, and even in the Black–Scholes model, the superreplication price of a call option is trivial as it equals the price of the underlying asset [2, 10, 20, 28, 40].

To overcome this dichotomy between transaction costs and frictionless markets, one needs to better understand the link between the law of asset price processes and their implications for contingent claim prices. This program, which, in frictionless markets, involves martingale measures, naturally leads to their transaction costs counterpart, a *consistent price system* [37]. This concept dates back to the seminal paper [21] and was further developed in the subsequent works [9, 22, 23, 25, 37].

A consistent price system (henceforth CPS) is essentially a shadow frictionless asset which admits a martingale measure and such that its price is always contained within the bid–ask spread of the original asset. The availability of CPS links the problems of no-arbitrage and superreplication to their frictionless counterparts, where solutions are well understood. Therefore, it is crucial to find general and easily testable conditions for their existence. In the previous literature [10, 28], asset prices were often assumed to be semimartingales, mainly for the sake of constructing classes of CPS. (See, however, [1, 7, 16, 20] where nonsemimartingale models are also considered.)

In the present paper, we study a simple condition on asset prices, namely *conditional full support*, which generates a large class of consistent price systems. In fact, all *natural* examples (which we can think of) enjoy this property. We study the problems of no-arbitrage and superreplication for asset prices driven by a continuous process and with constant proportional transaction costs.

Let $T > 0$ be a fixed time horizon. We consider a filtered probability space $(\Omega, \mathcal{F}, (\mathcal{F}_t)_{t \in [0,T]}, P)$, where the filtration $(\mathcal{F}_t)_{t \in [0,T]}$ satisfies the usual conditions of right-continuity and saturatedness with respect to $P$ as well as the conditions that $\mathcal{F}_T = \mathcal{F}$ and $\mathcal{F}_0$ is trivial. We introduce the notation $\mathbb{R}^d_{++} = (0, \infty)^d$. The set of $\mathbb{R}^d$-valued continuous functions on $[u, v]$ is denoted by $C[u, v]$, its subfamily of functions $f$ starting at $f(u) = x$ by $C_x[u, v]$. The class of $\mathbb{R}^d_{++}$-valued continuous functions on $[u, v]$ is denoted by $C^+[u, v]$. We endow these function spaces with the uniform norm topology.

When $x \in \mathbb{R}^d_{++}$, we write $C^+_x[u, v]$ for the set of $\mathbb{R}^d_{++}$-valued continuous functions on $[u, v]$ starting at $x$. The *support* of a $\mathbb{R}^d_{++}$-valued continuous process $S$ on $[u, v]$ is the smallest closed subset $A$ of $C^+[u, v]$ such that $P(S \in A) = 1$.

We say that the $\mathbb{R}^d_{++}$-valued process $S$ has *full support* if its support is the whole of $C^+_x[u, v]$, where $S_0 = x$. From now on, we shall use regular versions of conditional probabilities without further mention (see, e.g., page 439 of [3]).

We introduce the *conditional full support* condition which prescribes that from any given time on, the asset price path can continue arbitrarily close to any given path with positive conditional probability.

DEFINITION 1.1. A continuous $\mathbb{R}^d_{++}$-valued process $(S_t)_{t \in [0,T]}$ satisfies the *conditional full support* condition if, for all $t \in [0, T)$,

(CFS) $$\operatorname{supp} P(S|_{[t,T]} | \mathcal{F}_t) = C^+_{S_t}[t, T] \qquad \text{a.s.,}$$

where $P(S|_{[t,T]} | \mathcal{F}_t)$ denotes the $\mathcal{F}_t$-conditional distribution of the $C^+[t, T]$-valued random variable $S|_{[t,T]}$.

Our first main result shows that this condition implies the existence of consistent price systems.

THEOREM 1.2. *Let* $(S_t)_{t \in [0,T]}$ *be an* $\mathbb{R}^d_{++}$-*valued, continuous adapted process satisfying* (CFS). *$S$ then admits an $\varepsilon$-consistent pricing system for all $\varepsilon > 0$ (see Definition 2.1 below).*

The proof of this theorem is quite intuitive, at least in dimension $d = 1$: we show that any continuous price process satisfying the conditional full support condition (CFS) is arbitrarily close to the archetypal model of a "random walk with retirement," where martingale measures are characterized in terms of "retirement probabilities." Exploiting the flexibility of this construction, we discover a large class of consistent price systems and extend to our setting the "face-lifting" results proved by [10] and [4] for diffusion models. In particular, we may relax the Markov and even the semimartingale properties (see Section 4) which were required in the aforementioned papers; see also [20], Theorem 3, which applies to nonsemimartingale models, too.

THEOREM 1.3.  *Let $(S_t)_{t\in[0,T]}$ be an $\mathbb{R}_{++}$-valued continuous process satisfying the conditional full support assumption* (CFS) *and $g:\mathbb{R}_{++}\mapsto\mathbb{R}$ a lower semicontinuous function bounded from below.*

*The asymptotic superreplication price $p_0(g(S_T))$ (see Definition 2.18 below) of the European claim $g(S_T)$ is then given by*

$$p_0(g(S_T)) = \hat{g}(S_0),$$

*where $\hat{g}$ is the concave envelope of $g$.*

The $d$-dimensional extension of this result is left for future research. The proofs of these results help to explain the difference between transaction costs and frictionless markets as regards option pricing and hedging. Option pricers in the frictionless setting depend critically on the *fine* properties of the paths of the underlying price processes, such as quadratic variation, which are preserved only under equivalent changes of probabilities (see the recent study [1]; in the latest developments, even the condition (CFS) appears, [41]). Conceptually, such a dependence is rather questionable since these properties are hardly testable and are not preserved under arbitrarily small measurement errors. In contrast, accounting for transaction costs leads to results that are robust with respect to economically small misspecifications (i.e., within the bid–ask spread) and which depend only on the *coarse* properties of the model, such as the support of its law in $C_x^+[0,T]$. In this spirit, a consistent price system is the coarse version of a martingale measure.

Since our results are valid under the condition (CFS) of conditional full support, the crucial question is how to verify this assumption in concrete cases. Fortunately, this task can be accomplished by exploiting the literature on topological supports of stochastic processes, such as the classical results of [27] for Gaussian processes and [42] for diffusions.

As an application of our result, we show the existence of consistent price systems for fractional Brownian motion, which is a continuous process, but neither a Markov process, nor a semimartingale for Hurst parameter $H \neq 1/2$. This improves on a result in [17] where the absence of arbitrage was obtained under analogous assumptions, and paves the way for the study of optimization problems. In fact, with fractional Brownian motion, stochastic control methods are infeasible due to the lack of the Markov property, while the duality approach hinges on the availability of CPS.

The rest of this paper is organized as follows. In Section 2, we describe the model in detail, discuss some preliminaries on markets with transaction costs and prove Theorems 1.2 and 1.3 for a market with one risky asset, where simple proofs based on the Random Walk with Retirement are available. For the face-lifting result, we limit ourselves to this setting, where the fact that a one-dimensional random walk leaves any interval almost surely plays a crucial role in achieving a simple proof. Section 3 contains the proof of Theorem 1.2 in the general case of several assets. Here, we develop a similar idea as in the one-dimensional case, but need considerably more technique, since no simple analogy with retired random walks is available. Instead, we construct a conditional version of the classical Esscher transform.

As an application of our results, in Section 4, we show how to check the conditional full support assumption in the cases of Markov diffusions and fractional Brownian motion models. The Appendix contains some technical lemmas on conditional supports.

**2. One asset with proportional transaction costs.** In this section, we present the proofs of Theorems 1.2 and 1.3 for a market with one asset and with proportional transaction costs.

We assume that the bid and ask prices are given by $(1+\varepsilon)^{-1}S_t$ and $(1+\varepsilon)S_t$, respectively, where $(S_t)_{t\in[0,T]}$ is a continuous adapted process with strictly positive trajectories and $\varepsilon > 0$ is fixed.

We begin with the definition of a CPS, given here in the spirit of [21]:

DEFINITION 2.1. Let $\varepsilon > 0$. An $\varepsilon$-*consistent price system* is a pair $(\tilde{S}, \tilde{P})$ of a probability $\tilde{P}$ equivalent to $P$ and a $\tilde{P}$-martingale $\tilde{S}$ (adapted to $\mathcal{F}_t$) such that

$$(1) \qquad \frac{1}{1+\varepsilon} \le \frac{\tilde{S}_t^i}{S_t^i} \le 1+\varepsilon \qquad \text{a.s. for all } 1 \le i \le d \text{ and } t \in [0, T].$$

REMARK 2.2. We think of a consistent price system as a frictionless price $\tilde{S}$, offering better terms at all times, both for buying ($\tilde{S} \le S(1+\varepsilon)$) and selling [$\tilde{S} \ge S/(1+\varepsilon)$], than the process $S$ with $\varepsilon$ transaction costs.

It is then intuitively obvious (cf. Section 2.3) that any trading strategy executed at the price $\tilde{S}$ (and without transaction costs) yields a higher payoff than at the bid and ask of $S$ (paying transaction costs).

This observation has two immediate consequences. First, any no-arbitrage condition for the frictionless price $\tilde{S}$ extends to the bid–ask of $S$. In particular, if $S$ admits an $\varepsilon$-CPS for some $\varepsilon > 0$, then $S$ does not allow for free lunches (after paying $\varepsilon$-transaction costs) since $\tilde{S}$ has a martingale measure, by Definition 2.1.

Second, if some capital $x$ hedges a claim $X$ by trading on $S$ with transaction costs, it also hedges $X$ by trading on $\tilde{S}$ without frictions. Although this is also obvious at an intuitive level, the proof of Theorem 1.3 provides a formal argument. Then, as in the usual frictionless case, we obtain an inequality of the form

$$x \ge E_{\tilde{P}}[X],$$

where $\tilde{P} \sim P$ is a measure under which $\tilde{S}$ is a martingale.

Thus, the supremum of the right-hand side over the set of all consistent price systems yields a lower bound for the superreplication price. If this bound is large enough, one obtains Theorem 1.3, that is, trivial superreplication prices for options (asymptotically, as $\varepsilon \to 0$).

The previous discussions show that constructing consistent price systems is key to solving the no-arbitrage and superreplication problems under transaction costs by duality methods and in this section, we carry out this program in the one-dimensional case with constant proportional transaction costs. We begin by introducing the basic model of Random Walk with Retirement, which allows a large class of consistent price systems to be produced. In the following subsections, we employ this construction first to show the existence of consistent price systems (Theorem 1.2) and then to prove the face-lifting result for superreplication prices (Theorem 1.3).

2.1. *Random walk with retirement.* Consider a discrete-time filtered probability space $(\Omega, \mathcal{G}, (\mathcal{G}_n)_{n\geq 0}, P)$ such that $\mathcal{G}_0$ is trivial and $\bigvee_n \mathcal{G}_n = \mathcal{G}$.

DEFINITION 2.3. A *Random Walk with Retirement* is a process $(X_n)_{n\geq 0}$, adapted to $(\mathcal{G}_n)_{n\geq 0}$, of the form

(2)     $$X_n = X_0(1+\varepsilon)^{\sum_{i=1}^{n} R_i}, \qquad n \geq 1,$$

where $\varepsilon > 0$, $X_0 \in \mathbb{R}_{++}$ and the process $(R_n)_{n\geq 1}$ has values in $\{-1, 0, +1\}$ and satisfies:

  (i) $P(R_m = 0 \text{ for all } m \geq n | R_n = 0) = 1$ for $n \geq 1$;
  (ii) $P(R_n = x | \mathcal{G}_{n-1}) > 0$ on $\{R_{n-1} \neq 0\}$ for all $x \in \{-1, 0, +1\}$ and $n \geq 1$ (we set $\{R_0 \neq 0\} := \Omega$ as a convention);
  (iii) $P(R_n \neq 0 \text{ for all } n \geq 1) = 0$.

In plain English, a Random Walk with Retirement is just a random walk on the geometric grid $(X_0(1+\varepsilon)^k)_{k\in\mathbb{Z}}$, starting at $X_0$ and "retiring" at the a.s. finite stopping time $\rho = \min\{n \geq 1 : R_n = 0\}$. Note that the filtration $(\mathcal{G}_n)_{n\geq 0}$ is, in general, larger than the one generated by $X$.

In the following lemma, we describe the general form of a probability measure $Q \ll P$ such that $X$ is a $Q$-martingale. The martingale condition determines the relative weights of probabilities of upward and downward movements. By contrast, at each time, we may choose arbitrarily the conditional probability of retirement, denoted by $\alpha$.

LEMMA 2.4. *Let $(X_n)_{n\geq 0}$ be a Random Walk with Retirement, and $(\alpha_n)_{n\geq 1}$ a predictable (i.e., $\alpha_n$ is $\mathcal{G}_{n-1}$-measurable) process with values in*

$[0, 1]$. *If $\alpha$ satisfies*

$$\lim_{n \to \infty} E\left[\prod_{i=1}^{n}(1 - \alpha_i)\right] = 0, \tag{3}$$

*then there exists a (unique) probability $Q^\alpha$ on $\mathcal{G}$ such that:*

  (i) *$Q^\alpha$ is absolutely continuous with respect to $P$;*
  (ii) *$X$ is a $Q^\alpha$-martingale;*
  (iii) *$Q^\alpha(R_n = 0|\mathcal{G}_{n-1}) = \alpha_n$ a.s. on $\{R_{n-1} \neq 0\}$.*

*We have $Q \sim P$ iff $\alpha_n \in (0, 1)$ a.s. for $n \geq 1$.*

PROOF. Let $\alpha$ be given, satisfying (3). We explicitly construct the density $dQ^\alpha/dP$ and then check that it has the desired properties. We define the sequence of sets $(A_n)_{n \geq 0}$ by $A_0 = \varnothing$ and $A_n = \{R_n = 0\}$ for $n \geq 1$. By (i) and (iii) in Definition 2.3, this sequence a.s. increases to $\Omega$ and for $n \geq 1$, we define the $\mathcal{G}_n$-measurable random variable $Z_n$ as

$$Z_n = \begin{cases} \dfrac{\alpha_n 1_{\{R_n = 0\}}}{P(R_n = 0|\mathcal{G}_{n-1})} \\ \quad + \dfrac{\lambda_n 1_{\{R_n = -1\}}}{P(R_n = -1|\mathcal{G}_{n-1})} + \dfrac{\mu_n 1_{\{R_n = +1\}}}{P(R_n = +1|\mathcal{G}_{n-1})}, & \text{on } \Omega \setminus A_{n-1}, \\ 1, & \text{on } A_{n-1}, \end{cases} \tag{4}$$

where, on the set $\Omega \setminus A_{n-1}$, the nonnegative $\mathcal{G}_{n-1}$-measurable random variables $\lambda_n, \mu_n$ are uniquely chosen to satisfy the following two conditions:

$$E[Z_n|\mathcal{G}_{n-1}] = \alpha_n + \lambda_n + \mu_n = 1; \tag{5}$$

$$E[Z_n(1 + \varepsilon)^{R_n}|\mathcal{G}_{n-1}] = \alpha_n + (1 + \varepsilon)^{-1}\lambda_n + (1 + \varepsilon)\mu_n = 1. \tag{6}$$

We let $L_n = \prod_{i=1}^{n} Z_i$ and $L = \lim_{n \to \infty} L_n$. Since $A_n$ increases to $\Omega$ and $Z_m = 1$ on $A_m$ for $m > n$, it follows that $L = L_n$ on $A_n$ and $L > 0$ a.s.

$L$ defines a probability density if and only if $E[L] = 1$. We have, using (5),

$$E[L] = E\left[\lim_{n \to \infty}(L 1_{A_n})\right]$$
$$= \lim_{n \to \infty} E[L 1_{A_n}] = \lim_{n \to \infty} E[L_n 1_{A_n}] = 1 - \lim_{n \to \infty} E[L_n 1_{\Omega \setminus A_n}],$$

where the second equality follows by monotone convergence. Since $A_n$ increases to $\Omega$ and $Z_n$, by definition, satisfies $E[Z_n 1_{\Omega \setminus A_n}|\mathcal{G}_{n-1}] = 1_{\Omega \setminus A_{n-1}}(\lambda_n + \mu_n) = 1_{\Omega \setminus A_{n-1}}(1 - \alpha_n)$, iterating conditional expectations we obtain

$$E[L_n 1_{\Omega \setminus A_n}] = E\left[\prod_{i=1}^{n} Z_i \prod_{i=1}^{n} 1_{\Omega \setminus A_i}\right]$$

$$= E\left[\prod_{i=1}^{n}(Z_i 1_{\Omega \setminus A_i})\right] = E\left[E\left[\prod_{i=1}^{n}(Z_i 1_{\Omega \setminus A_i}) | \mathcal{G}_{n-1}\right]\right]$$

$$= E\left[(1-\alpha_n)1_{\Omega \setminus A_{n-1}}\prod_{i=1}^{n-1}(Z_i 1_{\Omega \setminus A_i})\right] = \cdots = E\left[\prod_{i=1}^{n}(1-\alpha_i)\right].$$

Thus, $E[L] = 1$ if

(7) $$0 = \lim_{n \to \infty} E\left[\prod_{i=1}^{n}(1-\alpha_i)\right],$$

which is (3). The probability measure $Q^\alpha(A) = E[L1_A]$ then satisfies (ii) and (iii) in view of (6) and (4).

To have $Q^\alpha \sim P$, it is necessary and sufficient that for all $n \geq 1$, $\alpha_n, \lambda_n, \mu_n > 0$ a.s. [because of (i), (ii) and (iii) of Definition 2.3], which is equivalent to $\alpha_n \in (0,1)$.  □

COROLLARY 2.5. *Any Random Walk with Retirement admits an equivalent martingale measure.*

PROOF. Apply the previous lemma to $\alpha_n \equiv 1/2$, for example.  □

The next lemma shows that, by choosing high probabilities of early retirement, one obtains an equivalent martingale measure with arbitrary integrability conditions. In particular, this implies the existence of equivalent martingale measures for which $X$ is uniformly integrable.

LEMMA 2.6. *Let $(X_n)_{n \geq 0}$ be a Random Walk with Retirement. Then, for any function $f : \mathbb{R}_{++} \mapsto \mathbb{R}$ and any $\varepsilon > 0$, there exists some $Q^\alpha \sim P$ as in Lemma 2.4 such that $E_{Q^\alpha}[\sup_{n \geq 0} f(X_n)] < \infty$.*

PROOF. Observe that the random variables $f(X_n)$ are supported on the double-sided sequence $(s_m)_{m \in \mathbb{Z}}$, where $s_m = f(x_m)$ and $x_m = X_0(1+\varepsilon)^m$. Observe, also, that we can limit ourselves to the case where

(8) $$s_m = s_{-m} \quad \text{and} \quad (s_m)_{m \geq 0} \text{ is increasing}$$

up to replacing the function $f$ with $\bar{f}(x_m) = \max_{|l| \leq |m|} f(x_l)$.

We set $M := \sup_{n \geq 0} f(X_n)$ and $Z_n = \sum_{i=1}^{n} R_i$ so that $X_n = X_0(1+\varepsilon)^{Z_n}$ and, by (8), we have that $\{M \geq s_m\} = \{\tau_m < \infty\}$, where $\tau_m = \min\{n : |Z_n| \geq m\}$.

Recalling (i) and (iii) of Definition 2.3, we know that, in fact, $M = \max_{n \geq 0} f(X_n)$. Using summation by parts, we have

$$E_{Q^\alpha}[M] = \sum_{m=0}^{\infty} s_m Q^\alpha(M = s_m) = s_0 + \sum_{m=1}^{\infty}(s_m - s_{m-1})Q^\alpha(M \geq s_m).$$

Also, the event that $M$ has reached the value $s_m$ is included in the event that the value $s_{m-1}$ has been reached by $M$ and retirement did not occur immediately after. Formally, we have

$$Q^\alpha(\tau_m < \infty | \tau_{m-1} < \infty) \leq Q^\alpha(R_{\tau_{m-1}+1} \neq 0 | \tau_{m-1} < \infty) = 1 - \alpha_{\tau_{m-1}} =: \delta_m,$$

whence the formula

$$(9) \qquad Q^\alpha(\tau_m < \infty) \leq \prod_{k=1}^m \delta_k =: \eta_m.$$

Thus, we obtain

$$(10) \qquad \sum_{m=1}^\infty (s_m - s_{m-1}) Q^\alpha(M \geq s_m) \leq \sum_{m=1}^\infty (s_m - s_{m-1}) \eta_m.$$

To have a convergent series, it is thus sufficient to choose $\eta_m$ small enough so that $(s_m - s_{m-1})\eta_n < 2^{-m}$. Since $\eta_m = \prod_{k=1}^m \delta_k$, this is equivalent to recursively choosing $\delta_m$ small or, equivalently, $\alpha_{\tau_{m-1}}$ close to one. To complete the definition of $\alpha$ outside the sequence of stopping times $(\tau_m)_{m \geq 1}$, we can choose an arbitrary value such as $\alpha = 1/2$ outside $\bigcup_{n=1}^\infty [\![\tau_n]\!]$. $\square$

The next lemma will be needed in Section 2.3 to characterize super-replication prices. It shows (by a suitable stopping time argument) that, given $S_0 \in \mathbb{R}_{++}$ and pairs of values $u, v$ satisfying $u < S_0 < v$, there exist $\varepsilon'$ (arbitrarily small), an $\varepsilon'$-random walk with retirement $X$ starting at $X_0 \in (S_0/(1+\varepsilon'), S_0(1+\varepsilon'))$ and $Q_\alpha \ll P$ such that $X_\infty$ is a.s. concentrated on $\{u, v\}$.

LEMMA 2.7. *Let the filtered probability space and $(R_n)_{n \geq 1}$ be as in Definition 2.3. Let $0 < u < S_0 < v$ be given. For all $\varepsilon$, there exist $\varepsilon' < \varepsilon$ and $Q^\alpha \ll P$ such that the $\varepsilon'$-Random Walk with Retirement starting at some $X_0 \in (S_0/(1+\varepsilon'), S_0(1+\varepsilon'))$ is such that $X$ is a $Q^\alpha$-martingale and*

$$(11) \qquad Q^\alpha(X_\infty = v) = \frac{X_0 - u}{v - u}, \qquad Q^\alpha(X_\infty = u) = \frac{v - X_0}{v - u}.$$

PROOF. There exists $\varepsilon' < \varepsilon$ such that for appropriate integers $j < 0 < k$ and $X_0 \in \mathbb{R}_{++}$, we have $u = X_0(1+\varepsilon')^j$, $v = X_0(1+\varepsilon')^k$ and $X_0 \in (S_0/(1+\varepsilon'), S_0(1+\varepsilon'))$. We define the stopping time

$$\tau = \min\{n : X_n = u \text{ or } X_n = v\}$$

and set $\alpha_i = 0$ if $i \leq \tau$ and $\alpha_i = 1$ otherwise, which trivially satisfies (3). We now apply the construction of Lemma 2.4. We get that $X$ is a bounded martingale under $Q^\alpha$, and the optional sampling theorem yields

$$X_0 = E_{Q^\alpha}[X_\tau] = E_{Q^\alpha}[X_\infty] = Q^\alpha(X_\infty = u)u + Q^\alpha(X_\infty = v)v.$$

This, combined with $Q^\alpha(X_\infty = u) + Q^\alpha(X_\infty = v) = 1$, implies the claim. $\square$

2.2. *Consistent price systems.* We now employ the previous construction to prove the existence of consistent price systems. We construct an increasing sequence of stopping times at which the process $S$ behaves like a "Random Walk with Retirement"; the conditional full support assumption is key to making this construction possible. Martingale measures for a Random Walk with Retirement are obtained by arbitrarily specifying the probability of retirement, as in Lemma 2.4 above. The $Q$-martingale $\tilde{S}$ is then defined as the continuous-time martingale determined by the terminal value of the random walk with retirement.

For technical reasons, we state a formally stronger version of the conditional full support condition in terms of stopping times.

DEFINITION 2.8. Let $\tau$ be a stopping time of the filtration $(\mathcal{F}_t)_{t \in [0,T]}$. Let us define $S_t := S_T$ for $t > T$ and let $\mu^\tau(\cdot, \omega)$ be (a regular version of) the $\mathcal{F}_\tau$-conditional law of the $C^+[0,T]$-valued random variable $(S_{\tau+t})_{t \in [0,T]}$.

We say that the *strong conditional full support condition* (SCFS) holds if, for each $[0,T]$-valued stopping time $\tau$ and for almost all $\omega \in \{\tau < T\}$, the following is true: for each path $f \in C^+_{S_\tau(\omega)}[0, T - \tau(\omega)]$ and for any $\eta > 0$, the $\eta$-tube around $f$ has positive $\mathcal{F}_\tau$-conditional probability, that is,

$$\mu^\tau(B_{f,\eta}(\omega), \omega) > 0,$$

where

$$B_{f,\eta} = \left\{ g \in C^+_{S_\tau(\omega)}[0,T] : \sup_{s \in [0, T-\tau(\omega)]} |f(s) - g(s)| < \eta \right\}.$$

In other words, this property means that for all $\tau$,

(SCFS)                 $\operatorname{supp} P(S|_{[\tau,T]}|\mathcal{F}_\tau) = C^+_{S_\tau}[\tau, T]$          a.s.,

that is, the conditional full support condition (CFS) also holds with respect to stopping times, while it was formulated in terms of deterministic times only in Definition 1.1 above.

The conditions (SCFS) and (CFS) are, in fact, equivalent. The precise formulation of this idea is somewhat technical, thus the proof of the next lemma is postponed to the Appendix.

LEMMA 2.9. *The conditional full support condition* (CFS) *implies the strong conditional full support condition* (SCFS), *hence they are equivalent.*

We now present the proof of Theorem 1.2 in dimension one and under the above (SCFS) hypothesis. In this case, the arguments are—hopefully—transparent and intuitive.

PROOF OF THEOREM 1.2 WHEN $d = 1$. We may suppose that $\varepsilon \in (0, 1)$. For any such $\varepsilon$, we associate to the process $(S_t)_{t \in [0,T]}$ a "random walk with retirement" as follows. We define the increasing sequence of stopping times

$$(12) \qquad \tau_0 = 0, \tau_{n+1} = \inf\left\{ t \geq \tau_n : \frac{S_t}{S_{\tau_n}} \notin ((1+\varepsilon)^{-1}, 1+\varepsilon) \right\} \wedge T.$$

For $n \geq 1$, we set

$$(13) \qquad R_n = \begin{cases} \mathrm{sign}\,(S_{\tau_n} - S_{\tau_{n-1}}), & \text{if } \tau_n < T, \\ 0, & \text{if } \tau_n = T. \end{cases}$$

Recall from the previous section the Random Walk with Retirement $(X_n)_{n \geq 0}$,

$$X_n = X_0 (1+\varepsilon)^{\sum_{i=1}^{n} R_i}$$

adapted to the filtration $(\mathcal{G}_n)_{n \geq 0}$, where $\mathcal{G}_n = \mathcal{F}_{\tau_n}$. To check the properties in Definition 2.3, observe that (i) is trivial, while (iii) follows from the continuity of paths. Furthermore, the (CFS) condition implies (ii) by Lemma A.1.

By Lemma 2.6, there exists some $Q^\alpha \sim P$ on $\mathcal{F} = \mathcal{G} := \bigvee_n \mathcal{G}_n$ such that

$$E_{Q^\alpha}\left[ \sup_{n \geq 0} X_n \right] < \infty.$$

Thus, $X$ is a uniformly integrable $(Q^\alpha, (\mathcal{G}_n)_{n \geq 0})$-martingale and is closed by its terminal value $X_\infty$. Define

$$\tilde{S}_t := E_{Q^\alpha}[X_\infty | \mathcal{F}_t], \qquad t \in [0, T].$$

Fix $0 \leq t \leq T$, define the random times $\sigma = \max\{\tau_n : \tau_n \leq t\}$ and $\tau = \min\{\tau_n : \tau_n > t\}$ and observe that $\tau$ is a stopping time. We have, by definition,

$$(1+\varepsilon)^{-1} \leq \frac{S_t}{S_\sigma}, \frac{S_\tau}{S_\sigma} \leq 1+\varepsilon \qquad \text{a.s. for all } t \in [0, T]$$

and we therefore obtain

$$(1+\varepsilon)^{-2} \leq \frac{S_\tau}{S_t} \leq (1+\varepsilon)^2 \qquad \text{a.s. for all } t \in [0, T].$$

By construction, $\tilde{S}_{\tau_n} = X_n$ and $S_{\tau_n} = X_n$ on $\{\tau_n < T\}$ for all $n \geq 0$. On $\{\tau_n = T\}$, we have the estimate

$$(1+\varepsilon)^{-1} \leq \frac{\tilde{S}_{\tau_n}}{S_{\tau_n}} \leq (1+\varepsilon)$$

for all $n \geq 0$.

The optional sampling theorem then implies that

$$\frac{\tilde{S}_t}{S_t} = \frac{E_{Q^\alpha}[\tilde{S}_\tau | \mathcal{F}_t]}{S_t} = E_{Q^\alpha}\left[ \frac{\tilde{S}_\tau}{S_\tau} \frac{S_\tau}{S_t} \Big| \mathcal{F}_t \right]$$

and therefore that

$$(1 + \varepsilon)^{-3} \leq \frac{\tilde{S}_t}{S_t} \leq (1 + \varepsilon)^3 \qquad \text{a.s. for all } t \in [0, T],$$

which completes the proof, up to the passage to a smaller $\varepsilon$.  $\square$

REMARK 2.10.  It is natural to consider a positive process $S$ and the corresponding random walk with retirement on a geometric grid. However, working on an arithmetic grid, the same argument yields the following result (stated without any reference to financial mathematics this time).

THEOREM 2.11.  *Let $(S_t)_{t \in [0,T]}$ be an adapted process on $(\Omega, \mathcal{F}, (\mathcal{F}_t)_{t \in [0,T]}, P)$ with continuous trajectories such that for all $t < T$,*

$$\operatorname{supp} P(S|_{[t,T]} | \mathcal{F}_t) = C_{S_t}[t, T],$$

*that is, the process has conditional full support on the space of (not only positive) paths. Then for all $\varepsilon > 0$, there exist $Q \sim P$ and a $Q$-martingale $(M_t)_{t \in [0,T]}$ such that*

$$|S_t - M_t| \leq \varepsilon \qquad \text{a.s. for all } t \in [0, T].$$

REMARK 2.12.  The construction in the proof of Theorem 1.2 bears similarities to the one appearing in [28]. Their Assumption 2.2 is weaker than the conditional full support condition. Indeed, they essentially require $P(R_n = \pm 1 | \mathcal{F}_{\tau_{n-1}}) > 0$ a.s. while (CFS) also ensures that $P(R_n = 0 | \mathcal{F}_{\tau_{n-1}}) > 0$ a.s. (see Lemma A.1). For the proof of Theorem 1.2, this latter property of "retirement with positive (conditional) probability" is indispensable.

To demonstrate this, we show (for simplicity) that Assumption 2.2 of [28] is not sufficient to obtain the conclusion of Theorem 2.11. Take $W_t$ to be standard Brownian motion and $\tau$ to be the first hitting time of 1. The process $S_t := W_{(\tan t) \wedge \tau}$, $t \in [0, \pi/2)$ with its natural filtration then satisfies Assumption 2.2 of [28] (indeed, it is a martingale on $[0, \pi/2)$). The limit $S_1$ exists and equals 1 almost surely. Take any $Q \sim P$ and any $Q$-martingale $M$ with $|S_T - M_T| < 1/3$. Then, necessarily, $M_t > 2/3$ a.s. for all $t$, which makes $|M_t - S_t| < 1/3$ a.s. impossible for $t < T$ since $S_t < 1/3$ with positive probability. The idea of this example comes from the remark after Lemma 1 of [26].

Another construction related to ours can be found in [20]; their Assumption 1 is also implied by (CFS).

2.3. *Superreplication.* We now prove the "face-lifting" Theorem 1.3, which characterizes superreplication prices of European options in terms of the concave envelopes of their payoffs evaluated at the current price of the underlying asset.

We first need some definitions and notation on trading strategies.

DEFINITION 2.13. A *trading strategy* is a predictable $\mathbb{R}$-valued process $\theta = (\theta_t)_{t \in [0,T]}$ such that $\theta_0 = \theta_T = 0$ and

$$(14) \qquad \sup_{0 \le t_0 \le \cdots \le t_n = T} \sum_{i=1}^{n} |\theta_{t_i} - \theta_{t_{i-1}}| < \infty \qquad \text{a.s.}$$

We may then define a predictable increasing process $\mathrm{Var}_s(\theta)$ satisfying a.s.

$$\mathrm{Var}_s(\theta) = \sup_{0 \le t_0 \le \cdots \le t_n = s} \sum_{i=1}^{n} |\theta_{t_i} - \theta_{t_{i-1}}| \qquad \text{for all } 0 \le s \le T.$$

We call $(\mathrm{Var}_s(\theta))_{0 \le s \le T}$ the total variation process of $\theta$.

DEFINITION 2.14. Given a trading strategy $\theta$, let us define

$$(15) \qquad V(\theta) := \int_0^T \theta_t \, dS_t - \varepsilon \int_0^T S_t \, d\mathrm{Var}_t(\theta)$$

and, for $t \in [0,T]$, introduce the random variables $V_t(\theta)$ as

$$(16) \qquad V_t(\theta) = V(\theta 1_{(0,t)})$$

so that $V(\theta) = V_T(\theta)$.

We call the process $\theta$ *M-admissible* for some $M \in \mathbb{R}$ if for all $t \in [0,T]$,

$$V_t(\theta) \ge -M$$

almost surely. The set of all such strategies is denoted by $\mathcal{A}_M$. We call $\mathcal{A} = \bigcup_{M>0} \mathcal{A}_M$, the class of *admissible strategies*.

REMARK 2.15. First, let us note that (14) implies that the integrals in (15) are a.s. well defined, in a pointwise Riemann–Stieltjes sense. Indeed, for the second integral, it suffices to observe that $S_t$ has continuous trajectories. As regards the first integral, it suffices to use partial integration and apply once more the previous argument.

REMARK 2.16. The random variable $V(\theta)$ clearly has the interpretation of the final gain or loss when applying the trading strategy $\theta$ of holding $\theta_t$ units of stock at time $t$: during the infinitesimal interval $[t, t+dt]$, the value of the position in stock (without considering transaction costs) changes by $\theta_t \, dS_t$, while one must pay $\varepsilon S_t \, d\mathrm{Var}_t(\theta)$ transaction costs. The condition $\theta_0 = \theta_T = 0$ corresponds to the requirement that we start and end without a position in the risky asset. Similarly, the term $1_{(0,t)}$ in the definition of $V_t(\theta)$ corresponds to the requirement that we take into account the liquidation cost at time $t$ to determine the value $V_t(\theta)$ of the portfolio at the given instant.

REMARK 2.17.   In the spirit of [39] or [44] in the frictionless case, and [22] in the case of transaction costs, we can also define a more general notion of admissibility. We say that the process $\theta$ is $M$-*admissible in the numéraire-free sense* if

$$V_t(\theta) \geq -M(1 + S_t).$$

In other words, a trading strategy $\theta$ is $M$-admissible in the numéraire-free sense if, by holding $M$ units of the bond *and* the stock, we make sure that the portfolio formed by the trading strategy $\theta$ and these two static positions has nonnegative liquidation value at all times $t \in [0, T]$.

Clearly, this definition leads to larger classes $\mathcal{A}_M^{nf}$ and $\mathcal{A}^{nf}$ of admissible trading strategies "in the numéraire-free sense." It can be checked that these strategies coincide with the admissible strategies in the sense of [5] when we restrict their framework to the present setting.

The name "numéraire-free" stems from the fact that this notion of admissibility is invariant under a change of numéraire between bond and stock.

It is shown in [44] for the frictionless case and in [18] for the case of transaction costs that the difference of the above two notions of admissibility corresponds to the difference of requiring the martingale or local martingale condition in Definition 1 above.

In the present paper, we formulate our results in terms of the notion of admissibility as given in Definition 2.14. The reader can check, however, that Theorem 1.3 still holds true if we choose in Definition 2.18 below the more general concept of *numéraire-free admissibility*.

DEFINITION 2.18.   In the setting of Definitions 2.13 and 2.14, consider a claim $X \in L^0(\mathcal{F}_T, P)$ and $\varepsilon > 0$. We define:

  (i) the *superreplication price*

$$p_\varepsilon(X) = \inf\{x : x + V(\theta) \geq X \text{ for some } \theta \in \mathcal{A}\};$$

  (ii) the *static superreplication price*

$$p_\varepsilon^{\mathrm{st}}(X) = \inf\{x : x + V(\alpha 1_{(0,T)}) \geq X \text{ for some } \alpha \in \mathbb{R}\};$$

(iii) the *asymptotic superreplication prices*

$$p_0(X) = \lim_{\varepsilon \downarrow 0} p_\varepsilon(X), \qquad p_0^{\text{st}}(X) = \lim_{\varepsilon \downarrow 0} p_\varepsilon^{\text{st}}(X).$$

REMARK 2.19. It is clear that $p_\varepsilon^{\text{st}}(X) \geq p_\varepsilon(X)$ and hence that $p_0^{\text{st}}(X) \geq p_0(X)$.

We can now proceed with the proof of Theorem 1.3. In view of the previous discussion, this theorem can be understood as follows: even considering the large class of admissible trading strategies $\mathcal{A}$, we cannot superreplicate a European contingent claim $X$ at a lower asymptotic ($\varepsilon \to 0$) cost than using only static hedges as in (ii) of Definition 2.18.

PROOF OF THEOREM 1.3. We first introduce the set of "absolutely continuous $\varepsilon$-consistent price systems,"

$$\mathcal{Z}_\varepsilon = \left\{ (\tilde{S}, \tilde{P}) : \tilde{P} \ll P, \ \tilde{S} \text{ is a } \tilde{P}\text{-martingale}, \ 1 - \varepsilon \leq \frac{\tilde{S}_t}{S_t} \leq 1 + \varepsilon, \ t \in [0, T] \right\}.$$

Note the minor deviation from (1): in this section, we take $1 - \varepsilon$ instead of $1/(1 + \varepsilon)$. It will become clear that this causes no problems since we let $\varepsilon$ tend to 0.

Consider the European claim $g(S_T)$, the initial capital $x$ and a superreplicating admissible strategy $\theta \in \mathcal{A}$ such that $x + V(\theta) \geq g(S_T)$ a.s.
If $(\tilde{S}, \tilde{P}) \in \mathcal{Z}_\varepsilon$, we obtain

$$\int_0^T \theta_t \, d\tilde{S}_t = - \int_0^T \tilde{S}_t \, d\theta_t$$

$$= V_T(\theta) + \varepsilon \int_0^T S_t \, d\text{Var}_t(\theta) + \int_0^T (S_t - \tilde{S}_t) \, d\theta_t \geq V_T(\theta)$$

and hence

$$g(S_T) \leq x + V(\theta) \leq x + \int_{[0,T]} \theta_t \, d\tilde{S}_t.$$

Since the right-hand side is a supermartingale by the admissibility of $\theta$ (it is uniformly bounded from below), it follows that

$$(17) \qquad p_\varepsilon(g(S_T)) \geq \sup_{\tilde{P} \in \mathcal{P}_\varepsilon} E_{\tilde{P}}[g(S_T)],$$

where

$$\mathcal{P}_\varepsilon := \{ \tilde{P} \ll P : \text{there exists } \tilde{S} \text{ such that } (\tilde{S}, \tilde{P}) \in \mathcal{Z}_\varepsilon \}.$$

Thus, to prove the theorem, it is sufficient to show that

$$\lim_{\varepsilon \downarrow 0} p_\varepsilon(g(S_T)) \geq \lim_{\varepsilon \downarrow 0} \sup_{\tilde{P} \in \mathcal{P}_\varepsilon} E_{\tilde{P}}[g(S_T)] \geq \hat{g}(S_0) \geq p_0^{\text{st}}(g(S_T)) \geq p_0(g(S_T)).$$

The first inequality is clear from (17), the second follows from Proposition 2.21 below, the third is a consequence of Proposition 2.20 below and the last one is trivial. By definition, the first quantity equals the last one, thus there are equalities throughout and the theorem is proved. □

PROPOSITION 2.20.   *Let* $g : \mathbb{R}_{++} \to \mathbb{R}$ *be a measurable function. Then,*

$$p_\varepsilon^{\mathrm{st}}(g(S_T)) \le \hat{g}(S_0) + J\varepsilon$$

*for a suitable constant* $J > 0$.

PROOF.   The case of $\hat{g}(S_0) = \infty$ being trivial, we may assume that $\hat{g}(S_0)$ is finite. By definition of the concave envelope,

$$\hat{g}(S_0) + \hat{g}'_+(S_0)(S_T - S_0) \ge g(S_T),$$

where $\hat{g}'_+$ is the right-hand derivative. Now, take $\beta \in \mathbb{R}$ satisfying $\beta - \varepsilon|\beta| = \hat{g}'_+(S_0)$. We then have

$$\hat{g}(S_0) + 2\varepsilon|\beta|S_0 + \beta(S_T - S_0) - \varepsilon|\beta|(S_T + S_0) \ge g(S_T).$$

Note that

$$p_\varepsilon^{\mathrm{st}}(g(S_T)) = \inf\{x : x + \alpha(S_T - S_0) - \varepsilon|\alpha|(S_0 + S_T) \ge g(S_T) \text{ for some } \alpha \in \mathbb{R}\},$$

which implies that $p_\varepsilon^{\mathrm{st}}(g(S_T)) \le \hat{g}(S_0) + 2\varepsilon|\beta|S_0$. □

PROPOSITION 2.21.   *Let* $g : \mathbb{R}_{++} \mapsto \mathbb{R}$ *be lower semicontinuous and bounded from below and denote by* $\hat{g}$ *its concave envelope. Then,*

$$\tag{18} \limsup_{\varepsilon \downarrow 0} \sup_{\tilde{P} \in \mathcal{P}_\varepsilon} E_{\tilde{P}}[g(S_T)] \ge \hat{g}(S_0).$$

PROOF.   Let us again suppose that $\hat{g}$ is finite-valued (the case $\hat{g} = \infty$ can be handled in a completely analogous manner).

It is enough to show that for all $\delta > 0$, and for all sufficiently small $\varepsilon > 0$ there exists $(\tilde{S}, Q) \in \mathcal{Z}_\varepsilon$ such that

$$\tag{19} E_{Q^\alpha}[g(S_T)] \ge \hat{g}(S_0) - \delta.$$

By definition of the concave envelope, there exist $0 < u < S_0 < v$ such that

$$\tag{20} g(u)\frac{v - S_0}{v - u} + g(v)\frac{S_0 - u}{v - u} > \hat{g}(S_0) - \frac{\delta}{3}$$

and, for $\varepsilon > 0$ small enough, we have

$$\tag{21} g(u)\frac{v - \tilde{S}_0}{v - u} + g(v)\frac{\tilde{S}_0 - u}{v - u} > \hat{g}(S_0) - \frac{\delta}{2}$$

for all $\tilde{S}_0 \in (S_0(1-\varepsilon), S_0(1+\varepsilon))$.

Since $g$ is lower semicontinuous, we can find, for $\varepsilon$ small enough, neighborhoods $U = (u/(1+\varepsilon), u(1+\varepsilon))$ and $V = (v/(1+\varepsilon), v(1+\varepsilon))$ such that $g(U) \geq g(u) - \delta/2$ and $g(V) \geq g(v) - \delta/2$.

Invoking Lemma 2.7 and recalling the construction in the proof of Theorem 1.2, there exist $Q \ll P$ and a $Q$-martingale $\tilde{S}$ such that

$$(22) \qquad Q(\tilde{S}_T = v) = \frac{\tilde{S}_0 - u}{v - u}, \qquad Q(\tilde{S}_T = u) = \frac{v - \tilde{S}_0}{v - u}$$

and

$$(23) \qquad |S_t - \tilde{S}_t| \leq \varepsilon' \qquad \text{a.s. for all } t \in [0, T]$$

holds for some $\varepsilon' < \varepsilon$ which is small enough to guarantee that $(\tilde{S}, Q) \in \mathcal{Z}^\varepsilon$.

Thus, we obtain

$$
\begin{aligned}
E_Q[g(S_T)] &= E_Q[g(S_T)1_{\{\tilde{S}_T = u\}}] + E_Q[g(S_T)1_{\{\tilde{S}_T = v\}}] \\
&\geq \left(g(u) - \frac{\delta}{2}\right)\frac{v - \tilde{S}_0}{v - u} + \left(g(v) - \frac{\delta}{2}\right)\frac{\tilde{S}_0 - u}{v - u} \\
&= g(u)\frac{v - \tilde{S}_0}{v - u} + g(v)\frac{\tilde{S}_0 - u}{v - u} - \frac{\delta}{2} \\
&> \hat{g}(S_0) - \delta,
\end{aligned}
$$

which is (19), as required. $\quad\square$

**3. The d-dimensional case.** In this section we prove the existence of consistent price systems in the multidimensional case. We consider a market $S = (S^0, S^1, \ldots, S^d)$ with one riskless asset $S^0$ and $d$ risky assets, based on a probability space $(\Omega, \mathcal{F}, (\mathcal{F}_t)_{t \in [0,T]}, P)$ satisfying the usual assumptions of right-continuity and saturatedness. We also assume that $\mathcal{F}_0$ is trivial and $\mathcal{F}_T = \mathcal{F}$. The riskless asset $S^0$ is used as numéraire and therefore assumed to be constantly equal to one. Each risky asset trades at bid and ask prices $S_t^i/(1+\varepsilon)$ and $S_t^i(1+\varepsilon)$, respectively, against the numéraire asset $S^0$.

In the higher-dimensional case, the random walks arising in our proof of Theorem 1.2 in the one-dimensional case are no longer supported on a finite grid, therefore we need a slightly different approach—which is similar in spirit—to construct a consistent price system.

The proof given below is based on a conditional version of the classical Esscher transform, which yields a recursive change of measure, without incurring too many technicalities related to measurable selections. This method was used by [33] to prove the main result of [11]; compare also (b) below to Theorem 3 of [19].

In the next lemma, $X$ represents the price increment over a given period, $\mathcal{G}$ and $\mathcal{H}$ the initial and final $\sigma$-algebras, respectively and $A$ the event where the process is already retired at the beginning of the period. For a subset $W \subset \mathbb{R}^d$, we denote by conv $W$ the convex hull of $W$, by int $W$ the interior of $W$ and by $\bar{B}_\delta$ the closed ball in $\mathbb{R}^d$ centered at the origin and with radius $\delta$.

LEMMA 3.1.   *Let $\mathcal{G} \subset \mathcal{H}$ be two $\sigma$-algebras, $A \in \mathcal{G}$, $X \in L^\infty(\Omega, \mathcal{H}, P; \mathbb{R}^d)$ and $\eta \in L^0(\Omega, \mathcal{G}, P; \mathbb{R}_{++})$ such that:*

(a)  $X = 0$ *on* $A$;
(b)  $0 \in \operatorname{int conv supp}(X|\mathcal{G})(\omega)$ *for almost all* $\omega \in \Omega \setminus A$;
(c)  $P(X = 0|\mathcal{G}) > 0$ *a.s.*

*There then exists $Z \in L^1(\mathcal{H}, \mathbb{R}_{++})$ such that almost surely:*

 (i)  $E[Z|\mathcal{G}] = 1$;
 (ii)  $E[ZX|\mathcal{G}] = 0$;
(iii)  $E[Z|X|^2|\mathcal{G}] \leq \eta$;
(iv)  $E[ZI_{\{X \neq 0\}}|\mathcal{G}] \leq \eta$.

PROOF.   We can restrict our attention to the set $\Omega \setminus A$ since setting $Z = 1$ on $A$ trivially satisfies (i)–(iv). We denote by $\mu(\omega, \cdot)$ the regular conditional law of $X$ with respect to $\mathcal{G}$.

As $X$ is bounded,

$$\theta \mapsto \phi(\theta, \omega) = \int e^{\theta \cdot X} \, d\mu(\omega, \cdot) < \infty \qquad \text{for all } \theta \in \mathbb{R}^d \text{ and a.e. } \omega.$$

Since $0 \in \operatorname{int conv supp}(X|\mathcal{G}) = \operatorname{int conv supp}\mu(\omega, \cdot)$, we can find a $\mathcal{G}$-measurable function $\delta: \Omega \mapsto \mathbb{R}_{++}$ such that $\operatorname{conv supp}(X|\mathcal{G}) = \operatorname{conv supp}\mu(\omega, \cdot) \supset \bar{B}_{\delta(\omega)}$. For instance, we can define $\delta$ as

$$1/\delta(\omega) = \min\{n : \bar{B}_{1/n} \subset \operatorname{conv supp}\mu(\omega, \cdot)\}.$$

Denoting by $\mathcal{S}$ the set of $\mathcal{G}$-measurable functions with values on the unit sphere of $\mathbb{R}^d$, we have that $\operatorname{conv supp}(v \cdot X|\mathcal{G}) \supset [-\delta, \delta]$ for any $v \in \mathcal{S}$, so $P(v \cdot X > \delta/2|\mathcal{G}) > 0$ a.s. In addition, we claim that (cf. Lemma 2.6 in [36])

$$\nu = \operatorname*{ess\,inf}_{v \in \mathcal{S}} P(v \cdot X > \delta/2|\mathcal{G}) > 0 \qquad \text{a.s.}$$

By contradiction, if $P(\nu = 0) > 0$, there would be a $\mathcal{G}$-measurable sequence $v_n \in \mathcal{S}$ achieving the essential infimum (as this set of conditional probabilities is easily seen to be directed downward) and by the compactness of the unit sphere, we could choose a $\mathcal{G}$-measurable random subsequence $n_k(\omega)$ such that $v_{n_k}$ converges almost surely to some $v' \in \mathcal{S}$ (see Lemma 2 of [24]). Then, $P(v' \cdot X \geq \delta/2 | \mathcal{G}) = 0$ with positive probability, which is absurd. Thus, we have, for all nonzero $\theta \in \mathbb{R}^d$,

$$\phi(\theta, \omega) = \int e^{\theta \cdot X} \, d\mu(\omega, \cdot) \geq e^{|\theta|\delta/2} \mu \left( \omega, \frac{\theta}{|\theta|} \cdot X > \delta/2 \right) \geq \nu e^{|\theta|\delta/2}$$

and therefore $\lim_{|\theta| \to \infty} \phi(\theta, \omega) = \infty$ almost surely. In particular, the function $\phi(\cdot, \omega)$ admits a unique [by the strict convexity of $\theta \mapsto \phi(\theta, \omega)$] minimum $\theta^*(\omega)$ which solves $\nabla \phi(\theta^*(\omega), \omega) = 0$ and is hence $\mathcal{G}$-measurable. Dominated convergence then implies that

$$0 = \nabla \phi(\theta^*(\omega), \omega) = \int X e^{\theta^*(\omega) \cdot X} \, d\mu(\omega, \cdot) = E[X e^{\theta^* \cdot X} | \mathcal{G}]$$

and $Z' = e^{\theta^* \cdot X} / E[e^{\theta^* \cdot X} | \mathcal{G}]$ satisfies properties (i)–(ii). To further obtain (iii) and (iv), we rescale, conditionally on $\mathcal{G}$, the relative weight of the events $\{X = 0\}$ and $\{X \neq 0\}$, which does not affect the expectation of $X$. Formally, we denote by $Y := P(X = 0 | \mathcal{G})$ and set $Z = \lambda Z' + \mu 1_{\{X=0\}}$, where the $\mathcal{G}$-measurable positive functions $\lambda$ and $\mu$ will be chosen so as to satisfy (i), that is

$$(24) \qquad\qquad\qquad \lambda + \mu Y = 1.$$

It is clear that $E[Z'|X|^2 | \mathcal{G}] < \infty$ almost surely. Hence, we can choose a $\mathcal{G}$-measurable $\lambda > 0$ so small that not only both $\lambda E[Z'|X|^2 | \mathcal{G}] \leq \eta$ and $\lambda \leq \eta$ hold, but also (24) is satisfied for an appropriate $\mathcal{G}$-measurable $\mu > 0$. This completes the proof.  $\square$

PROOF OF THEOREM 1.2 IN THE $d$-DIMENSIONAL CASE.   We define the increasing sequence of stopping times,

$$\tau_0 = 0,$$

$$\tau_{n+1} = \inf \left\{ t \geq \tau_n : \frac{S_t^i}{S_{\tau_n}^i} \notin ((1+\varepsilon)^{-1}, 1+\varepsilon) \text{ for some } 1 \leq i \leq d \right\} \wedge T.$$

Path continuity implies that $\tau_n = T$ a.s. for all $n \geq \bar{n}(\omega)$, hence the events $A_n = \{\tau_n = T\}$ increase to $\Omega$ as $n$ grows to infinity. For $n \geq 1$, we set

$$\Delta_n := (S_{\tau_n} - S_{\tau_{n-1}}) 1_{\Omega \setminus A_n}.$$

Obviously, $\mathrm{supp}(\Delta_n | \mathcal{F}_{\tau_{n-1}}) = \{0\}$ on $A_{n-1}$. Assumption (CFS) implies that $0 \in \mathrm{int\,conv\,supp}(\Delta_n | \mathcal{F}_{\tau_{n-1}})$ and $P(\Delta_n = 0 | \mathcal{F}_{\tau_{n-1}}) > 0$ on $\Omega \setminus A_{n-1}$; this fact is shown in Lemma A.2 below.

For $n \geq 1$, we now recursively apply Lemma 3.1 to

$$A = A_{n-1}, \qquad \mathcal{G} = \mathcal{F}_{\tau_{n-1}}, \qquad \mathcal{H} = \mathcal{F}_{\tau_n} \qquad X = \Delta_n, \qquad \eta := 2^{-n},$$

obtaining a sequence of strictly positive $\mathcal{F}_{\tau_n}$-measurable random variables $(Z_n)_{n \geq 1}$ such that, almost surely:

  (i) $E[Z_n|\mathcal{F}_{\tau_{n-1}}] = 1$;
  (ii) $E[\Delta_n Z_n|\mathcal{F}_{\tau_{n-1}}] = 0$;
  (iii) $E[Z_n|\Delta_n|^2|\mathcal{F}_{\tau_{n-1}}] \leq 2^{-n}$;
  (iv) $E[Z_n 1_{\{\Delta_n \neq 0\}}|\mathcal{F}_{\tau_{n-1}}] \leq 2^{-n}$.

From the construction, it follows that $Z_n = 1$ on $A_m$ for all $n \geq m + 1$. Therefore, letting $L_n = \prod_{i=1}^{n} Z_i$, we have

$$L = \lim_{n\to\infty} L_n = \prod_{n=1}^{\infty} Z_n > 0 \qquad \text{a.s.}$$

because $A_n \uparrow \Omega$.

    Using (iv) above, we obtain

$$E[L] = E\left[L \lim_{n\to\infty} 1_{A_n}\right] = \lim_{n\to\infty} E[L 1_{A_n}]$$

$$= 1 - \lim_{n\to\infty} E[L_n 1_{\Omega\setminus A_n}] \geq 1 - \lim_{n\to\infty} 2^{-n} = 1.$$

Hence, the density $L$ induces the probability measure $Q(A) := E[L 1_A]$ on $\mathcal{F}$. We now define the discrete-time process $(M_n)_{n\geq 0}$ as

$$M_n = S_0 + \sum_{l=1}^{n} \Delta_l.$$

By construction, $M$ is a $Q$-martingale with respect to the filtration $(\mathcal{F}_{\tau_n})_{n\geq 0}$. We have $M_n = S_{\tau_n}$ a.s. on $\{\tau_n < T\}$, while on $\{\tau_n = T\}$, a.s.,

$$(1+\varepsilon)^{-1} \leq \frac{M_n^i}{S_{\tau_n}^i} = \frac{S_{\tau_{n-1}}^i}{S_T^i} \leq 1+\varepsilon \qquad \text{for all } t \in [0,T],\ 1 \leq i \leq d,\ n \geq 0.$$

Hence, we always have that, a.s.

$$(25) \qquad (1+\varepsilon)^{-1} \leq \frac{M_n^i}{S_{\tau_n}^i} \leq 1+\varepsilon \qquad \text{for all } t \in [0,T],\ 1 \leq i \leq d,\ n \geq 0.$$

We claim that $M$ is uniformly integrable, in fact, that it is bounded in $L^2$. Indeed, we have, from (iii), that

$$E_Q[|M_n|^2] = |S_0|^2 + \sum_{l=1}^{n} E_Q|\Delta_l|^2$$

$$= |S_0|^2 + \sum_{l=1}^{n} E[L_{l-1} E[Z_l|\Delta_l|^2|\mathcal{F}_{\tau_{l-1}}]]$$

$$\leq |S_0|^2 + \sum_{l=0}^{\infty} 2^{-l} < \infty.$$

Letting $\sigma = \max\{\tau_n : \tau_n \le t\}$ and $\tau = \min\{\tau_n : \tau_n > t\}$, since $t \in [\sigma, t] \cap [\sigma, \tau]$, we have

$$(1+\varepsilon)^{-1} \le \frac{S_t^i}{S_\sigma^i}, \frac{S_\tau^i}{S_\sigma^i} \le 1+\varepsilon \qquad \text{a.s. for all } t \in [0,T], \ 1 \le i \le d,$$

and therefore

$$(26) \qquad (1+\varepsilon)^{-2} \le \frac{S_\tau^i}{S_t^i} \le (1+\varepsilon)^2 \qquad \text{a.s. for all } t \in [0,T], \ 1 \le i \le d.$$

We define the martingale $\tilde{S}_t = E_Q[M_\infty | \mathcal{F}_t], \ t \in [0,T]$, which satisfies $\tilde{S}_{\tau_n} = M_n$ for all $n \ge 0$ by construction and by the uniform integrability of $M$. Hence, combining (25) and (26) with

$$\frac{\tilde{S}_\tau^i}{S_t^i} = \frac{\tilde{S}_\tau^i}{S_\tau^i} \frac{S_\tau^i}{S_t^i},$$

it follows that

$$(1+\varepsilon)^{-3} \le \frac{\tilde{S}_\tau^i}{S_t^i} \le (1+\varepsilon)^3 \qquad \text{a.s. for all } t \in [0,T], \ 1 \le i \le d.$$

Since $\tilde{S}_t = E[\tilde{S}_\tau | \mathcal{F}_t]$ by optional sampling, the theorem follows up to the passage to a smaller $\varepsilon$. $\quad\square$

**4. Applications.** To illustrate the scope of our results, we now present some important classes of models where the conditions of Theorem 1.2 can be checked.

4.1. *Markov processes.*

EXAMPLE 4.1. Consider a (homogeneous) a.s. continuous Markov process $S_t, t \in [0,T]$, with state space $\mathbb{R}_{++}^d$ such that for all $x \in \mathbb{R}_{++}^d$, starting from $S_0 = x$, the process has full support on $C_x^+[0,T]$. The Markov property then immediately implies the conditional full support condition

$$\text{supp } P(S|_{[v,T]} | \mathcal{F}_v) = \text{supp } P(S|_{[v,T]} | S_v) = C_{S_v}^+[v,T], \qquad 0 \le v \le T.$$

Methods to show that a diffusion process has full support can be found in the seminal work of [42] and in [32] (page 340).

4.2. *Fractional Brownian motion.* We now turn to models based on fractional Brownian motion (FBM). Recall that $(X_t)_{t\ge0}$ is FBM with Hurst parameter $0 < H < 1$ if it is a centered Gaussian process with continuous sample paths and covariance function

$$\Gamma(t,s) = \tfrac{1}{2}(t^{2H} + s^{2H} - |t-s|^{2H}).$$

If $H = 1/2$, we recover the standard Brownian motion. For a thorough treatment of FBM, we refer the reader to the monograph [31].

Models of asset prices based on fractional Brownian motion have long attracted the interest of researchers for their properties of long-range dependence [8, 29, 30, 43]. However, in a frictionless setting, it turns out that these models lead to arbitrage opportunities [6, 12, 14, 34, 35, 38] and therefore cannot be meaningfully employed for studying optimal investment and derivatives pricing.

In this paper, we show how this situation is completely different as soon as arbitrarily small transaction costs are introduced. There then exist consistent price systems and hence the duality theory and hedging theorems of [25] and [5] apply.

The next result improves on Proposition 5.1 of [17] and follows from a similar argument.

PROPOSITION 4.2. *Let $S_t = \exp\{\sigma X_t + f_t\}$, where $X_t$ is FBM with parameter $0 < H < 1$ and $f_t$ is a deterministic continuous function. $(S_t)_{t \in [0,T]}$ then satisfies the conditional full support condition (CFS) with respect to its (right-continuous and saturated) natural filtration.*

PROOF. Let us fix $v \in [0, T]$. It is enough to prove that the conditional law $P(X|_{[v,T]}|\mathcal{F}_v)$ has full support on $C_{X_v}([v, T], \mathbb{R})$ almost surely.

From the representation of Corollary 3.1 in [13], we know that, for some square-integrable kernel $K_H(t, s)$, one has

$$(27) \qquad X_t = \int_0^t K_H(t, s) \, dW_s,$$

for some Brownian motion $(W_t)_{t \in [0,T]}$ generating the same filtration as $(X_t)_{t \in [0,T]}$.

It is easily seen, by directly calculating the conditional joint characteristic function of finite-dimensional distributions of $X$, that for any $v \in [0, T]$, the process $(X_t)_{t \in [v,T]}$ is Gaussian, conditionally on $\mathcal{F}_v$. Its conditional expectation and conditional covariance function are given by

$$
\begin{aligned}
(28) \qquad & c_t := E[X_t | \mathcal{F}_v] = \int_0^v K_H(t, s) \, dW_s, && t \geq v, \\
& \tilde{\Gamma}(t, s) := \mathrm{cov}_{\mathcal{F}_v}(X_t, X_s) = \int_v^{t \wedge s} K_H(t, u) K_H(s, u) \, du, && t, s \geq v.
\end{aligned}
$$

Observe that $\tilde{\Gamma}(t, s)$ does not depend on $\omega$. Hence, for almost all $\omega$, the law of $(X_t)_{t \in [v,T]}$ conditional on $\mathcal{F}_v$ is equal to the law of $Y_t + c_t(\omega)$, where $(Y_t)_{t \in [v,T]}$ is a centered Gaussian process with continuous paths on $[v, T]$ and with covariance function $\tilde{\Gamma}$. Thus, recalling the kernel representation (27), it suffices to prove that the centered Gaussian process

$$Y_t := \int_v^t K_H(t,s)\,dW_s, \qquad t \in [v,T],$$

has full support on $C_0([v,T],\mathbb{R})$.

Theorem 3 in [27] states that the topological support of a continuous Gaussian process $(Y_t)_{t\in[v,T]}$ is equal to the norm closure of its reproducing kernel Hilbert space, defined by

$$\mathbb{H} := \left\{ f \in C_0([v,T],\mathbb{R}) : f(t) = \int_v^t K_H(t,s)g(s)\,ds, \text{ for some } g \in L^2[v,T] \right\}.$$

Thus, it is sufficient to show that $\mathbb{H}$ is norm-dense in $C_0([v,T],\mathbb{R})$.

To achieve this, we need to recall the Liouville fractional integral operator for any $f \in L^1[a,b]$ and $\gamma > 0$,

$$(I_{a+}^\gamma f)(t) := \frac{1}{\Gamma(\gamma)} \int_a^t f(s)(t-s)^{\gamma-1}\,ds, \qquad a \le t \le b,$$

and to introduce the kernel operator $K_H$,

$$(K_H f)(t) := \int_0^t K_H(t,s)f(s)\,ds, \qquad f \in L^2[0,T],\ t \in [0,T].$$

We first treat the case $H < 1/2$. In this case, we have, by [13], Theorem 2.1, that

$$K_H f = I_{0+}^{2H}(s^{1/2-H} I_{0+}^{1/2-H}(s^{H-1/2}f(s))).$$

From now on, we assume that all functions on a subset of $\mathbb{R}$ are extended to the whole real line by setting them to be 0 outside their domain of definition. In the case $v = 0$, we could perform the same calculations as in [17], pages 578–579. For general $v$, the argument needs to be split into two steps.

LEMMA 4.3. *If $f \in C_0[v,T]$, then $L_1 f \in C_0[v,T]$, where*

$$(L_1 f)(t) = (I_{0+}^{1/2-H}(s^{H-1/2}f(s)))(t).$$

*Moreover, $L_1 : C_0[v,T] \to C_0[v,T]$ is continuous and has a dense range (with respect to the uniform norm).*

PROOF. Clearly, $L_1 f$ is a continuous function and $(L_1 f)(0) = 0$. The operator is continuous by the estimate

$$\|L_1 f - L_1 g\|_\infty \le v^{H-1/2} \int_0^T (T-s)^{-H-1/2}\,ds \|f - g\|_\infty.$$

Recall the identity for $a, b > 0$,

$$\int_0^t (t-u)^{a-1} u^{b-1}\,du = C(a,b)t^{a+b-1},$$

where $C(a,b) \neq 0$ is a constant. Defining, for a fixed $\alpha > 0$,

$$g(s) := 1_{[v,T]} \frac{(s-v)^\alpha}{s^{H-1/2}},$$

we obtain, for $t \in [v,T]$,

$$(L_1 g)(t) = \int_v^t (t-s)^{-H-1/2} g(s) s^{H-1/2} \, ds = \int_v^t (t-s)^{-H-1/2}(s-v)^\alpha \, ds$$

$$= \int_0^{t-v} u^\alpha (t-v-u)^{-H-1/2} \, du = C(\alpha+1, 1/2 - H)(t-v)^{\alpha - H + 1/2}.$$

Varying $\alpha$, we find that $(t-v)^n \in \operatorname{Im}(L_1)$ for $n \geq 1$ and the Stone–Weierstrass theorem guarantees that $\operatorname{Im}(L_1)$ is dense in $C_0[v,T]$.  $\square$

LEMMA 4.4.  *If* $f \in C_0[v,T]$, *then* $L_2 f \in C_0[v,T]$, *where*

$$(L_2 f)(t) = (I_{0+}^{2H}(s^{1/2-H} f(s)))(t)$$

*and* $L_2 \colon C_0[v,T] \to C_0[v,T]$ *is continuous and has a dense range.*

PROOF.  The same argument applies, but this time we use the estimation

$$\|L_1 f - L_1 g\|_\infty \leq T^{1/2-H} \int_0^T (T-s)^{2H-1} \, ds \|f-g\|_\infty$$

and the function

$$g(s) := 1_{[v,T]} \frac{(s-v)^\alpha}{s^{1/2-H}}. \qquad \square$$

Since the restriction of $K_H$ to $C_0[v,T]$ is exactly $L_2 \circ L_1$, we may conclude that $K_H \colon C_0[v,T] \to C_0[v,T]$ has a dense range and, a fortiori, $\mathbb{H}$ is norm-dense in $C_0[v,T]$.

In the case $H \geq 1/2$, a similar representation holds [13], Theorem 2.1,

$$K_H f = I_{0+}^1(s^{H-1/2} I_{0+}^{H-1/2}(s^{1/2-H} f)),$$

and the same argument carries over.  $\square$

4.3. *Processes with smooth trajectories.*  In frictionless market models one is accustomed to processes with rather irregular paths. As already discussed in the introduction, such "fine" properties do not matter when transaction costs are present. One may construct arbitrage-free models with smooth trajectories using, for instance, the next lemma.

LEMMA 4.5.    *Let $X$ be a continuous adapted process on $(\Omega, \mathcal{F}, (\mathcal{F}_t)_{t \in [0,T]}, P)$, satisfying the following conditional full support condition:*

$$(29) \qquad \operatorname{supp} P(X|_{[t,T]}|\mathcal{F}_t) = C_{X_t}[t,T] \qquad \text{for all } 0 \le t < T.$$

*Then, the process*

$$Y_t := \int_0^t X_s \, ds, \qquad t \in [0,T],$$

*also satisfies (29).*

PROOF (Sketch).    Let us suppose that the statement is false. Then, for some $t < T$, $B \in \mathcal{F}_t$ with $P(B) > 0$ and for all $\omega \in B$, there exist $f_\omega \in C_{Y_t}[t,T]$ and $\eta > 0$ such that

$$(30) \qquad \nu_Y^t\left(\left\{g: \sup_{s \in [t,T]} |f_\omega(s) - g(s)| < \eta\right\}, \omega\right) = 0,$$

where $\nu_Y^t$ is the $\mathcal{F}_t$-conditional (regular) law of $Y|_{[t,T]}$. Using measurable selection and the fact that smooth functions are dense in $C[t,T]$, we may suppose that $f_\omega(s)$ is continuously differentiable (in $s$) and $\mathcal{F}_t$-measurable (in $\omega$). We may also suppose that $f'_\omega(t) = X_t$ almost surely. By the hypothesis on $X$,

$$\nu_X^t\left(\left\{h: \sup_{s \in [t,T]} |f'_\omega(s) - h(s)| < \eta/(T-t)\right\}, \omega\right) > 0$$

for almost all $\omega \in B$, in other words,

$$P\left(\sup_{s \in [t,T]} |X_s - f'(s)| < \eta/(T-t)|\mathcal{F}_t\right) > 0 \qquad \text{a.s. on } B.$$

It follows [by integration of $X_s$ and $f'(s)$] that

$$P\left(\sup_{s \in [t,T]} |Y_s - Y_t - f(s) + f(t)| < \eta|\mathcal{F}_t\right) > 0$$

a.s. on $B$. Recall that $Y_t = f(t)$, which contradicts (30).    □

In view of the above lemma, taking any $X$ satisfying (29) (such as fractional Brownian motion; see the previous subsection) and defining $S_t := \exp(Y_t)$, $t \in [0,T]$, we obtain a process with the property (CFS) and with continuously differentiable trajectories. Iterating the integration, it is possible to obtain processes with even smoother trajectories.

## APPENDIX

We collect here a few arguments of a rather technical nature. First, we show the equivalence of (SCFS) and (CFS) by contradiction: if there is an open set of paths with $\mathcal{F}_\tau$-conditional probability 0 on a set of positive measure, then we can find a deterministic time $q$ "close enough" to $\tau$ such that the same phenomenon arises, which is absurd, by (CFS).

PROOF OF LEMMA 2.9. By contradiction, suppose that, for some stopping time $\tau$, there exists $\{\tau < T\} \supset A \in \mathcal{F}_\tau$ with $P(A) > 0$ such that, for almost all $\omega \in A$, there exist $f_\omega \in C_1^+[0, T]$ and $\eta_\omega > 0$ such that $\mu^\tau(B(\omega), \omega) = 0$, where

$$B(\omega) := \left\{ g \in C^+_{S_{\tau(\omega)}(\omega)}[0, T] : \sup_{s \in [0, T - \tau(\omega)]} |g(s) - f_\omega(s) S_{\tau(\omega)}(\omega)| \le \eta_\omega \right\}.$$

The measurable selection theorem (see Sections III. 44–45 of [15]) enables us to choose $\omega \to (f_\omega, \eta_\omega)$ in an $\mathcal{F}_\tau$-measurable way. Set $K_\omega := \|f_\omega\|_\infty \|1/f_\omega\|_\infty$.

Now define, for each $q \in Q := \mathbb{Q} \cap [0, T)$,

$$A_q := A \cap \{q \ge \tau\} \cap \left\{ \sup_{s \in [\tau, q]} |f(s - \tau) S_\tau - S_s| \le \frac{\eta}{2K} \right\} \in \mathcal{F}_q.$$

For almost all fixed $\omega$, the functions $f_\omega(s)$, and $S_s(\omega)$ are continuous in $s$. As $f_\omega(0) S_{\tau(\omega)}(\omega) = S_{\tau(\omega)}(\omega)$, the functions $s \to f_\omega(s - \tau) S_{\tau(\omega)}(\omega)$ and $s \to S_s(\omega)$ remain closer to each other than $\eta_\omega / 2K_\omega$ on some interval $[\tau(\omega), q(\omega)]$, where $q(\omega) > \tau(\omega)$ may be chosen to be rational.

This shows that $A = \bigcup_{q \in Q} A_q$, hence we may fix $q$ such that $P(A_q) > 0$. Define

$$C_q := \left\{ \sup_{s \in [q, T]} |f(s - \tau) S_\tau - S_s| \le \eta \right\},$$

$$G_q := \left\{ \sup_{s \in [q, T]} \left| \frac{S_q f(s - \tau)}{f(q - \tau)} - S_s \right| \le \frac{\eta}{2} \right\} \cap \{q \ge \tau\},$$

$$H := \left\{ \sup_{s \in [\tau, T]} |f(s - \tau) S_\tau - S_s| \le \eta \right\}.$$

Note that on $A_q \cap G_q$, we have, for $q \le s \le T$,

$$|f(s - \tau) S_\tau - S_s| \le \left| f(s - \tau) S_\tau - \frac{S_q f(s - \tau)}{f(q - \tau)} \right| + \left| \frac{S_q f(s - \tau)}{f(q - \tau)} - S_s \right|$$

$$\le \frac{f(s - \tau)}{f(q - \tau)} \frac{\eta}{2K} + \frac{\eta}{2} \le \eta,$$

hence, $A_q \cap G_q \subset A_q \cap C_q$. As $A_q \cap C_q \subset A \cap H$ and

$$P(A \cap H) = E[1_A E[1_H | \mathcal{F}_\tau]] = \int_\Omega 1_A(\omega) \mu^\tau(B(\omega), \omega) \, dP(\omega) = 0,$$

we obtain

$$0 = E[1_{A_q} 1_{G_q}] = E[1_{A_q} E[1_{G_q} | \mathcal{F}_q]],$$

which is a contradiction as $P(G_q | \mathcal{F}_q) > 0$ almost surely by the (CFS) condition. $\square$

LEMMA A.1. *Let $S$ be an $\mathbb{R}_{++}$-valued continuous process satisfying* (CFS) *and let $R_n$ be defined by* (13). *Then, $P(R_{n+1} = z | \mathcal{F}_{\tau_n}) > 0$ a.s. on $\{\tau_n < T\}$ for $z = 0, \pm 1$ and $n \geq 0$.*

PROOF. We write $\tau$ instead of $\tau_n$ and set $D := \{\tau < T\}$. Define the (random) function

$$f(t) := S_\tau \left( 1 + \frac{2t\varepsilon}{T - \tau} \right)^z, \qquad t \in [0, T - \tau],$$

$$f(t) := f(T - \tau), \qquad t > T - \tau.$$

Define $\eta := S_\tau \varepsilon / 2$. Lemma 2.9 implies that $\mu^\tau(B(\omega), \omega) > 0$ for almost all $\omega \in D$, where $B(\omega)$ is the following set of paths:

$$B(\omega) := \left\{ g \in C^+_{S_{\tau(\omega)}(\omega)}[0, T] : \sup_{s \in [0, T - \tau(\omega)]} |f_\omega(s) - g(s)| \leq \eta(\omega) \right\}.$$

First, take $z = 0$. For $\omega \in D$, paths in $B(\omega)$ hit neither $S_\tau(\omega)(1 + \varepsilon)$ nor $S_\tau(\omega)(1 + \varepsilon)^{-1}$ on $[0, T - \tau(\omega)]$, so

$$P(S_{\tau_{n+1}} = S_{\tau_n} | \mathcal{F}_\tau) = P(\tau_{n+1} = T | \mathcal{F}_\tau)$$

$$\geq P\left( \sup_{s \in [\tau, T]} |f(s - \tau) - S_s| \leq \eta | \mathcal{F}_\tau \right)$$

$$= \mu^\tau(B(\omega), \omega) > 0 \quad \text{on } D$$

and we are done.

If $z = \pm 1$, then for $\omega \in D$, each path in $B(\omega)$ attains $S_\tau(\omega)(1 + \varepsilon)^z$ on $[0, \frac{3}{4}(T - \tau(\omega))]$ without attaining $S_\tau(\omega)(1 + \varepsilon)^{-z}$, so

$$P(S_{\tau_{n+1}} = S_{\tau_n} (1 + \varepsilon)^z | \mathcal{F}_\tau)$$

$$\geq P\left( \sup_{s \in [\tau, T]} |f(s - \tau) - S_s| \leq \eta | \mathcal{F}_\tau \right) = \mu^\tau(B(\omega), \omega) > 0 \qquad \text{on } D,$$

as claimed. $\square$

The next lemma is the multidimensional counterpart of Lemma A.1. It shows that, under (CFS), there is a positive probability of hitting each face of the surface of the "bid–ask cube" ($F$ in the proof below). In the one-asset case, it coincides with two points, while, in general, it has $d-1$ dimensions. The idea of the proof is to see that paths which move linearly (either up or down) along a single coordinate must hit their respective face with positive probability. Hence, there is a neighborhood on each face which has positive conditional probability, thus the support must contain some point from each of these faces and therefore the interior of its convex hull contains the origin.

LEMMA A.2.  *Suppose that the assumptions of Theorem 1.2 hold true and recall the notation in its proof in Section 3. On the set $\Omega \setminus A_{n-1}$, we have, almost surely,*

$$(31) \qquad 0 \in \operatorname{int} \operatorname{conv} \operatorname{supp}(\Delta_n | \mathcal{F}_{\tau_{n-1}}) \quad \text{and} \quad P(\Delta_n = 0 | \mathcal{F}_{\tau_{n-1}}) > 0.$$

PROOF.   Take $D := \Omega \setminus A_{n-1}$. We will write $\tau$ for $\tau_{n-1}$. Let $F$ be the (random) cube with edges $(S_\tau(1+\varepsilon)^{l(1)}, \ldots, S_\tau(1+\varepsilon)^{l(d)}), l \in \{\pm 1\}^d$. Also, define its faces $F_{iz} = \operatorname{ri}\{x \in F : x^i = (1+\varepsilon)^z\}$, $z \in \{\pm 1\}, 1 \le i \le d$, where ri stands for "relative interior."

Introduce the random functions

$$f_{iz}^i(t) := S_\tau^i \left(1 + \frac{2t\varepsilon}{T-\tau}\right)^z, \qquad t \in [0, T-\tau],$$

$$f_{iz}^j(t) := S_\tau^j, \qquad\qquad\qquad j \neq i, t \in [0, T-\tau],$$

$$f_{iz}(t) := f_{iz}(T-\tau), \qquad\qquad t > T-\tau,$$

for $z \in \{\pm 1\}$ and $1 \le i \le d$. Set $\eta^{iz} := \varepsilon \min_i S_\tau^i / 2$.

We obtain as in the previous proof, that on $D$, with positive $\mathcal{F}_\tau$-probability, the trajectory $S_t^i$ attains $S_\tau^i(1+\varepsilon)^z$ on $[\tau, \tau + 3/4(T-\tau)]$, while the other coordinates $S_t^j, j \neq i$, remain in the intervals $(S_\tau^j(1+\varepsilon)^{-1}, S^j(1+\varepsilon))$. That is,

$$P(S_{\tau_n} - S_\tau = \Delta_n \in F_{iz} | \mathcal{F}_\tau) > 0 \qquad \text{on } D,$$

for each $z, i$, which means that the $\mathcal{F}_\tau$-conditional support of $\Delta_n$ contains some point of each face of the "bid–ask cube" $F$, hence its convex hull contains 0 in its interior.

The proof of $P(\Delta_n = 0 | \mathcal{F}_\tau) > 0$ on $D$ also follows the argument of the previous lemma.   □

**Acknowledgements.**   The authors thank Yuri M. Kabanov and Ch. Stricker for their preprint "On martingale selectors of cone-valued processes" and two anonymous referees for their reports. We are also grateful to A. Cherny for his preprint "Brownian moving averages have conditional full support."

P. Guasoni
Department of Mathematics and Statistics
Boston University
111 Cummington St
Boston, Massachusetts 02215
USA
E-mail: guasoni@bu.edu

M. Rásonyi
Computer and Automation Institute
  of the Hungarian Academy of Sciences
P.O. Box 63 1518 Budapest
Hungary
and
Vienna University of Technology
Research Unit of Financial
  and Actuarial Mathematics,
Wiedner Hauptstrasse 8-10/105-1
A-1040 Vienna
Austria
E-mail: rasonyi@sztaki.hu

W. Schachermayer
Vienna University of Technology
Research Unit of Financial
  and Actuarial Mathematics
Wiedner Hauptstrasse 8-10/105-1
A-1040 Vienna
Austria
E-mail: wschach@fam.tuwien.ac.at